\documentclass[aps,preprint]{revtex4}
\usepackage{amsfonts}
\usepackage{amsmath}
\usepackage{amssymb}
\usepackage{color}
\usepackage{ulem}
\usepackage{graphicx}
\usepackage{epstopdf}
\usepackage{float}
\usepackage{subcaption}

\begin{document}
\title{Lyapunov exponents and phase transition of charged Ads black hole in quintessence}
\author{Xiaobo Guo$^{a}$}
\email{guoxiaobo@foxmail.com}

\author{Rui Yang$^{b}$}
\email{yangrui7@stu.scu.edu.cn}

\author{Yizhi Liang$^{b}$}
\email{liangyizhi@stu.scu.edu.cn} 

\author{Jun Tao$^{b}$}
\email{taojun@scu.edu.cn}
\affiliation{$^{a}$ Department of Statistics, Shanghai University of Finance and Economics Zhejiang College, Jinhua, 321015, China}
\affiliation{$^{b}$College of Physics, Sichuan University, Chengdu, 610065, China}

\begin{abstract}
This study investigates the phase transitions of RN-AdS black holes immersed in a quintessence field, employing Lyapunov exponents as a dynamical probe to characterize the thermodynamics of black hole. Incorporating quintessence dark energy into the RN-AdS framework, we find that the Lyapunov exponents for null and timelike geodesics display diminished chaotic behavior with increasing normalization factor of the quintessence field. This feature introduces a finite cutoff to the Lyapunov exponent of unstable circular photon orbits, setting it apart from RN-AdS black hole. At phase transition points, both the free energy and Lyapunov exponents display multivalued branches, reflecting the coexistence of distinct black hole phases. Furthermore, the discontinuity in the Lyapunov exponent can serve as an order parameter with a critical exponent of $1/2$ near the critical point, consistent with the thermodynamic criticality of van der Waals fluids.. These work suggest that Lyapunov exponents provide a framework for probing the thermodynamics of black holes.
\end{abstract}
\maketitle
\tableofcontents

\section{Introduction}
The exploration of black hole thermodynamics greatly changed our understanding of the fundamental connections among gravity, quantum mechanics, and statistical physics. Hawking's theorem demonstrate that the total horizon area of black holes cannot decrease over time \cite{Hawking:1971tu}, and Bekenstein conjectures that black hole entropy is proportional to its horizon area \cite{Bekenstein:1973ur}. The discovery of Hawking radiation further strengthened this framework, linking the surface gravity of a black hole with its temperature \cite{Hawking:1974rv}. Subsequently, Bardeen, Carter, and Hawking formalized the four laws of black hole thermodynamics, establishing a analogy with classical thermodynamics \cite{Bardeen:1973gs}.

The study of black holes in anti-de Sitter (AdS) spacetimes has introduced further depth to this field, through the perspective of the AdS/CFT correspondence \cite{Maldacena:1997re}. An important result in this context is the Hawking-Page phase transition, which describes a transition between thermal AdS space and Schwarzschild-AdS black holes \cite{Hawking:1982dh}. For charged AdS black holes, such as the Reissner-Nordström-AdS (RN-AdS) configuration, thermodynamic phase transitions exhibit similarities to the van der Waals liquid-gas system, characterized by a first-order phase transition between small and large black hole phases \cite{Chamblin:1999tk, Chamblin:1999hg}. These transitions have been probed using thermodynamic quantities like free energy and heat capacity \cite{Guo:2021enm, He:2016fiz}.

The accelerating expansion of the Universe \cite{SupernovaSearchTeam:1998fmf, SupernovaCosmologyProject:1998vns}, attributed to dark energy with negative pressure \cite{Caldwell:1997ii}, has motivated modifications to gravitational theories. Quintessence as scalar fields characterized by the equation of state parameter $\omega \in (-1,  -1/3)$ \cite{Rodrigue:2018dlg, Thomas:2012zzc, Copeland:2006wr, Caldwell:1997ii, Ratra:1987rm}, which causes the acceleration \cite{SupernovaCosmologyProject:1998vns}. When coupled to black holes, these fields induce asymptotically non-flat spacetime structures that alter thermodynamic behavior \cite{Kiselev:2002dx}. Kiselev's solution demonstrated that quintessence surrounds black holes as an anisotropic fluid, modified critical exponents in $P$-$V$ phase diagrams, and a state-dependent Hawking temperature $T_H(\omega)$ via surface gravity renormalization \cite{Li:2014ixn}. Recent studies confirm that for $\omega < -1/2$, quintessence destabilizes small charged AdS black holes by shifting the spinodal line in $T$-$S$ planes \cite{Tan:2024sgv, Ghaderi:2024ihe}. For related discussions on black holes enveloped by quintessence, see Refs \cite{Minazzoli:2012md, Chen:2008ra, Wei:2011za, Thomas:2012zzc, Tharanath:2013jt, Toshmatov:2017btq, Ghaderi:2016dpi, Ma:2016arz, Xu:2016jod, Saleh:2017vui, Ghosh:2017cuq, Ghaffarnejad:2018exz, Xu:2018nar, Wu:2018meo, Toledo:2019amt, Ndongmo:2019ywh, Chabab:2020ejk, Wu:2020tmz}.

 The dynamical behavior of black holes can be effectively analyzed through a framework combining thermodynamic developments and chaos theory. The Lyapunov exponent is a key indicator of chaos which quantifies the rate of divergence or convergence of nearby trajectories  \cite{LYAPUNOV1992The, Cardoso:2008bp}. In black hole physics, Lyapunov exponents have been employed to investigate the stability of particle orbits \cite{Cardoso:2008bp, Wang:2016wcj}. Recent studies reveal the connection between Lyapunov exponents and black hole phase transitions, where a discontinuous change in the exponent can signal a phase transition \cite{Guo:2022kio, Liu:2014gvf}. This relationship provides a novel dynamical probe into the phase structure of black holes. In this paper we will discuss the relationship between thermodynamic phase transition of charged AdS black holes in quintessence (RN-qAdS black holes) and the Lyapunov exponent. 

The structure of this paper is as follows: In Section ~\ref{sec:thermodynamics}, we discussion the thermodynamic properties of charged AdS black holes in quintessence. In section ~\ref{sec:Lyapunov}, we analyzes the phase transitions in the parameter space, deriving the Lyapunov exponents for timelike and null geodesics and their correlations with phase transitions. The critical behavior near the transition point, defining an order parameter based on the Lyapunov exponent are investigated in Section~\ref{sec:critical_exponent}. Finally we give a conclusion of the whole paper in section~\ref{sec:concl}. 

\section{Thermodynamics of RN-qAdS black hole}
\label{sec:thermodynamics}

In this section, we explore the thermodynamic properties and phase transitions of Reissner-Nordström-AdS black holes surrounded by quintessence in four-dimensional spacetime. The action for this system is given by \cite{Minazzoli:2012md, Hong:2019yiz}
    \begin{align}
    S = \int \mathrm{d}^4 x \sqrt{-g} \left[ \frac{1}{16 \pi G} (R - 2 \Lambda - F_{\mu \nu} F^{\mu \nu}) + \mathcal{L}q \right],
    \end{align}
where $R$ is the Ricci scalar, $\Lambda$ is the cosmological constant, $F_{\mu \nu}$ is the electromagnetic tensor and $\mathcal{L}_q$ represents the contribution from quintessence, modeled as $\mathcal{L}_q = -\rho \left[ 1 + \omega \ln (\rho / \rho_0) \right]$ with an integration constant $\rho_0$ \cite{Minazzoli:2012md}.  The cosmological constant is related to the AdS radius by
    \begin{align}
    \Lambda = -\frac{(d-1)(d-2)}{2 l^2}=-\frac{3}{l^2}.
    \end{align}
$l$ is the AdS radius. Under static spherical symmetry, the metric takes the form \cite{Kiselev:2002dx, Alfaia:2021cnk}
    \begin{align}
    ds^2 = f(\tilde{r}) dt^2 - \frac{1}{f(\tilde{r})} d\tilde{r}^2 - \tilde{r}^2 d\Omega^2,
    \end{align}
with the metric function
    \begin{align}
    f(\tilde{r}) = 1 - \frac{2\tilde{M}}{\tilde{r}} + \frac{\tilde{Q}^2}{\tilde{r}^2} + \frac{\tilde{r}^2}{l^2} - \frac{\tilde{b}}{\tilde{r}^{3\omega + 1}}.\label{eq:f(r)}
    \end{align}
$\tilde{M}$ is the black hole mass, $\tilde{Q}$ is the charge. $\tilde{b}$ as a fixed parameter characterizing the quintessence energy density which is a positive normalization factor and there exists an upper bound for $\tilde{b}$, if the Hawking-Page phase transitions can occur.

From Eq. \eqref{eq:f(r)}, the event horizon radius $\tilde{r}_+$ is the largest root of $f(\tilde{r}_+) = 0$, allowing us to express the RN-qAdS black hole's mass $\tilde{M}$ as
    \begin{align}
    \tilde{M} = \frac{1}{2} ( -\tilde{b} \tilde{r}_+^{-3\omega} + \frac{\tilde{r}_+^3}{l^2} + \tilde{r}_+ + \frac{\tilde{Q}^2}{\tilde{r}_+} ). \label{eq:M_r+}
    \end{align}
To avoid naked singularity, $\tilde{r}_{+}$ must be positive, which sets the lower bound on $\tilde{M}$ as $\tilde{M}>\tilde{Q}$ \cite{Yan:2021uzw}.

The analog of thermodynamic expression presented by the first law of black hole thermodynamics can be given as
    \begin{align}
    \mathrm{d} \tilde{M}= \tilde{T} \mathrm{~d} S +\tilde{\Phi} \mathrm{d} \tilde{Q}, \label{eq:dm}
    \end{align}
where $\tilde{T}$, and $\tilde{\Phi} = \tilde{Q} / \tilde{r}_+$ are the Hawking temperature and electric potential at the event horizon of the RN-qAdS black hole. The Hawking temperature at the event horizon, derived from the surface gravity can be rewritten in terms of $\tilde{r}_+$ as
    \begin{align}
    \tilde{T} = -\frac{\tilde{Q}^2}{4 \pi \tilde{r}_+^3} + \frac{1}{4 \pi \tilde{r}_+} + \frac{3 \tilde{r}_+}{4 \pi l^2} + \frac{3 \tilde{b} \omega \tilde{r}_+^{-2 - 3 \omega} }{4 \pi}. \label{eq:T}
    \end{align}
The Bekenstein-Hawking entropy of the black hole is given by 
    \begin{align}
    S=\frac{A}{4} = \pi \tilde{r}_+^2. \label{eq:S}
    \end{align}
Through evaluation of the Euclidean action within the semiclassical framework, the free energy of the black hole satisfying
    \begin{align}
    \tilde{F}=\tilde{M}-\tilde{T} S=\frac{\tilde{Q}^2}{2 \tilde{r}_{+}}+\frac{\tilde{r}_{+}}{4}+\frac{1}{2 l^2} \tilde{r}_{+}^3-(\frac{3}{4}+ \frac{\tilde{b}}{2}+\frac{3 \tilde{b} \omega}{4} )\tilde{r}_{+}^{-3 \omega}. \label{eq:F}
    \end{align}
For dimensional consistency, we use scaled quantities:
    \begin{align}
    r_+ = \tilde{r}_+ / l, \quad Q = \tilde{Q} / l, \quad b = \tilde{b} / l^{3\omega + 1}, \quad T = \tilde{T} l, \quad M = \tilde{M} / l, \quad F = \tilde{F} / l. \label{eq:scaled}
    \end{align}

We can use the Hawing temperature expressed in Eq. \eqref{eq:T} to find the critical points for phase transitions
    \begin{align}
        \frac{\partial T}{\partial r_+} = 0, \quad \frac{\partial^2 T}{\partial r_+^2} = 0, \label{eq:partial_T}
    \end{align}
yielding
    \begin{align}
        -3 - \frac{3 Q^2}{r_+^4} + \frac{1}{r_+^2} + 3 b\omega (2 + 3 \omega) r_+^{-3(1 + \omega)}  = 0, \\
         -12 Q^2 r_+^{3 \omega} + 2 r_+^{2 + 3 \omega} + 9 b\omega (2 + 5 \omega + 3 \omega^2) r_+  = 0. \label{eq:critical_cond}
    \end{align}
Solving these gives critical $Q_c$ and $b_c$ for fixed $r_+$
    \begin{align}
        Q_c &= \frac{\sqrt{r_+^2 (1 + 3 \omega - 9 r_+^2 (1 + \omega))}}{\sqrt{-3 + 9 \omega}}, \label{eq:Q_c}\\
        b_c &= -\frac{2 r_+^{1 - 3 \omega} (-1 + 6 r_+^2)}{3 \omega (-2 + 3 \omega + 9 \omega^2)}. \label{eq:b_c}
    \end{align}

\begin{figure}[h]
    \centering
    \includegraphics[width=0.6\textwidth]{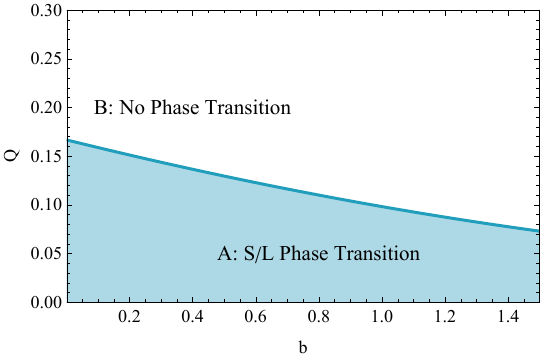}
    \caption{The critical curve in the $Q$-$b$ parameter space for $\omega = -1/2$, separating regions with and without phase transitions.}
    \label{fig:b_vs_q}
    \end{figure}

The critical curve in the normalization factor and  charge ($b-Q$) space is shown in Fig.~\ref{fig:b_vs_q} for $\omega = -1/2$. 
Compared to RN-AdS black holes, the quintessence term $-\tilde{b} / \tilde{r}^{3\omega + 1}$ introduces additional complexity to the parameter space, potentially leading to richer phase structures under certain conditions. The $b-Q$ space divides into two regions. Region A where $b < b_c$ or $Q < Q_c$ along with a small/large (S/L) phase transition that exhibits van der Waals-like behavior characterized by first-order transitions. Region B where $b > b_c$ or $Q > Q_c$ exists a stable single phase without phase transition.

\begin{figure}[htbp]
\centering
\begin{subfigure}{0.43\textwidth}
\includegraphics[width=\textwidth]{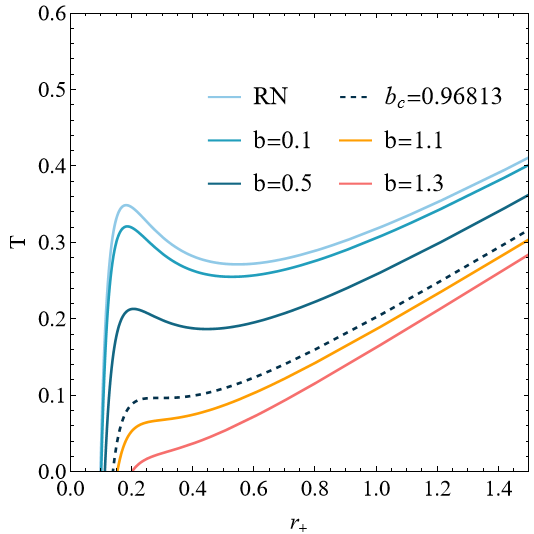}
\caption{$T-r_+$ for different $b$.}
\label{fig:Tvsr+}
\end{subfigure}
\hfill
\begin{subfigure}{0.50\textwidth}
\includegraphics[width=\textwidth]{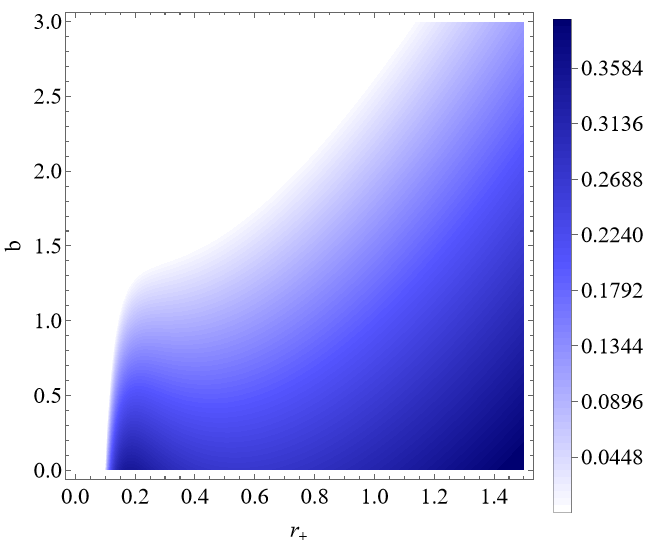}
\caption{$T\left(r_{+}, b\right)$}
\label{fig:Tvsb,r+}
\end{subfigure}
\caption{Effect of the quintessence parameter $b$ on Hawking temperature $T$.}
\label{fig:influence_T}
\end{figure}

Furthermore, we choose the charge $Q = 0.1$, critical values are 
\begin{align}
    r_{+,c} \approx 0.291531, \quad b_c \approx 0.968133, \quad T_c \approx 0.096415.
\end{align}
The variation of temperature with the horizon radius is illustrated in Fig.~\ref{fig:Tvsr+}. For $b < b_c$, the $T$-$r_+$ curve has two extrema, indicating S/L transitions. For $b > b_c$, it is monotonic, confirming no transition in this area. At $b = b_c$ , an inflection point signals a second-order transition. The temperature $T$ is presented as a contour plot in the $b$ and $r_{+}$ plane. The boundary where the positive Hawking temperature shifts leftward as quintessence parameter increases, indicating that larger normalization factor permit higher cut-off values of $r_{+}$.

\begin{figure}[htbp]
\centering
\begin{subfigure}{0.47\textwidth}
\includegraphics[width=\textwidth]{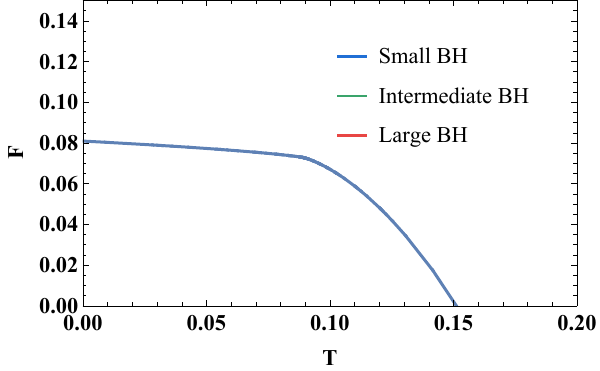}
\caption{$b=1>b_c$.}
\label{fig:FvsT_b1}
\end{subfigure}
\hfill
\begin{subfigure}{0.47\textwidth}
\includegraphics[width=\textwidth]{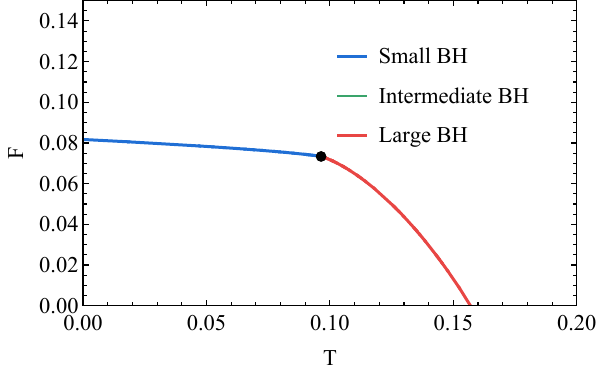}
\caption{$b=b_c$.}
\label{fig:FvsT_bc}
\end{subfigure}

\vspace{1cm}

\begin{subfigure}{0.47\textwidth}
\includegraphics[width=\textwidth]{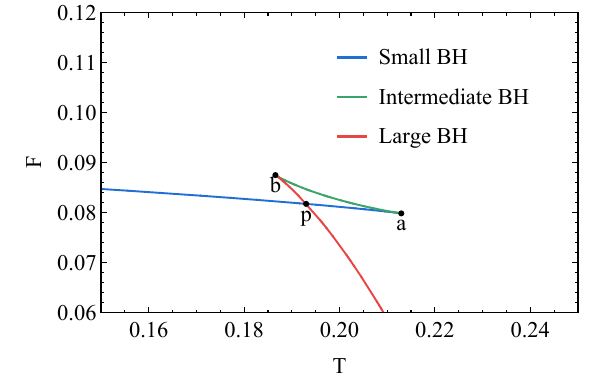}
\caption{$b=0.5<b_c$.}
\label{fig:FvsT_b0.5}
\end{subfigure}
\hfill
\begin{subfigure}{0.47\textwidth}
\includegraphics[width=\textwidth]{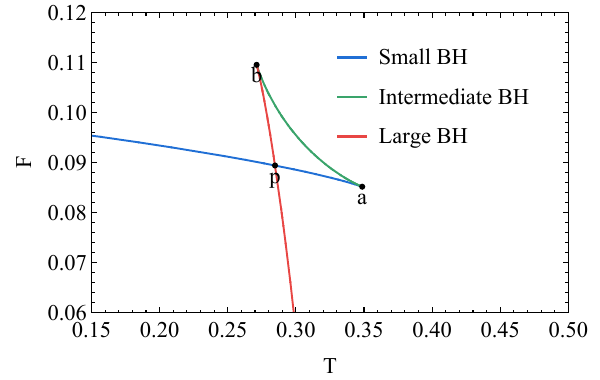}
\caption{$b=0<b_c$.}
\label{fig:FvsT_b0}
\end{subfigure}
\caption{F-T for different $b$ values, with  $Q=0.1$ and $b_c \approx 0.968133$. }
\label{fig:FvsT_b_various}
\end{figure}

To reveal the thermodynamic stability and transition properties of these black holes, we plot free energy $F$ as a function of Hawking temperature $T$  for various $b$ values in Fig.~\ref{fig:FvsT_b_various}. For $b < b_c$, a typical swallow tail pattern forms with three branches linked to small, intermediate, and large black holes, and resulting a van der Waals-like first-order phase transition from small to large black holes at the point $p$, as seen in the In Fig.~\ref{fig:FvsT_b0.5}. A same swallow tail structure emerged in the case $b$ approaches zero which resembles the Reissner-Nordström configuration as showing in Fig.~\ref{fig:FvsT_b0}. In Fig.~\ref{fig:FvsT_bc} for $b = b_c$, the black hole transit from small to large at a specific point, involving a second-order phase transition. In Fig.~\ref{fig:FvsT_b1} for $b > b_c$, free energy reduces steadily with rising temperature, showing no small to large phase transition.

\section{Lyapunov Exponents of RN-qAdS black hole}
\label{sec:Lyapunov}

The Lyapunov exponent acts as a crucial indicator for the sensitivity and complexity in dynamical systems, revealing whether trajectories diverge or converge. This approach applies to studying divergence rates near black holes. Lyapunov exponents have previously probed small/large phase transitions in RN-AdS black hole\cite{Guo:2022kio}. We extend this method to quintessence. In particular, we investigate the connection between the Lyapunov exponents of massive and massless particles in unstable circular orbits nearest to the event horizon of RN-qAdS black hole and the associated thermodynamic phase transitions of these black holes.
\subsection{Lyapunov Exponents and Thermodynamics for Null Geodesics}
\label{sub:null}

\begin{figure}[htbp]
\centering
\begin{subfigure}{0.47\textwidth}
\includegraphics[width=\textwidth]{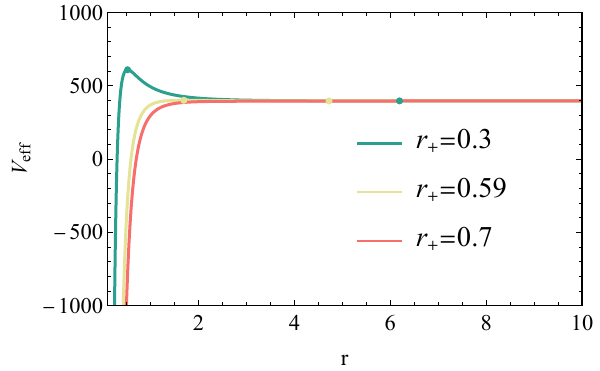}
\caption{$V_{\text{eff}}(r) - r$ for varying $r_+$.}
\label{fig:Veff_null}
\end{subfigure}
\hfill
\begin{subfigure}{0.47\textwidth}
\includegraphics[width=\textwidth]{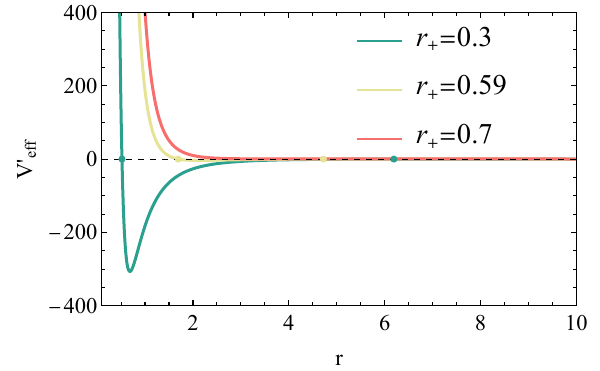}
\caption{$V_{\text{eff}}^\prime(r) - r$ for varying $r_+$.}
\label{fig:Veff_prime_null}
\end{subfigure}
\caption{Effective potential $V_{\text{eff}}(r)$ and its derivative $V_{\text{eff}}^\prime(r)$ for different $r_+$.}
\label{fig:Veff_analysis(null)}
\end{figure}

Photon trajectories follow null geodesics in the equatorial plane ($\theta=\pi / 2$), parameterized by the affine parameter and the Hamiltonian reads
    \begin{align}
    -2 \mathcal{H} = \frac{\dot{r}^2}{f(r)} + L \dot{\varphi} - E \dot{t} = 0,
    \end{align}
where where $L=\tilde{L}/l$ is scaled with angular momentum and $E$ aslo denotes scaled with conserved energy. The effective potential governing radial motion is
    \begin{align}
    V_{\text{eff}}(r) = f(r) \left( \frac{L^2}{r^2} - \frac{E^2}{f(r)} \right).
    \end{align}
The circular orbits occur where $V_{\text{eff}}^\prime(r)= 0$ and the Unstable condition satisfying $V^{\prime \prime}(r_c) < 0$, resulting the Lyapunov exponent \cite{Lyu:2023sih}
    \begin{align}
    \lambda = \sqrt{-\frac{r_c^2 f(r_c)}{2 L^2} V^{\prime \prime}(r_c)},
    \label{eq:lambda_null}
    \end{align}
$r_c$ represents the radius at the unstable circular orbit. The Lyapunov exponent is closely related to the effective potential . To understand the RN-qAdS black hole dynamics better, we analyze the effective potential $V_{\text{eff}}(r)$ and its derivative $V_{\text{eff}}^\prime(r)$ with respect to the radius with $Q=0.1$, $L=20$ in Fig.~\ref{fig:Veff_analysis(null)}. The effective potential is a complex function influenced by parameters including the horizon radius $r_+$, the quintessence normalization factors $b$, and the charge $Q$. 
In Fig.~\ref{fig:Veff_null}, the effective potential $V_{\text{eff}}(r)$ exhibits a maximum outside the event horizon, corresponding to the unstable photon sphere and approaching constant at large $r$. For smaller $r_+$ such as $0.3$, the maximum is higher and located at smaller $r$, resulting in a steeper drop post-maximum. As $r_+$ increases, the maximum lowers and shifts outward, with the overall curve flattening. This indicates that larger horizons yield shallower potentials, reducing the barrier for instability and affecting orbital dynamics. Markedly, for sufficiently large $r_+$, the maximum vanishes, eliminating unstable circular orbits. 

The derivative $V_{\text{eff}}'(r)$ is displayed in Fig.~\ref{fig:Veff_prime_null}, starting positive near the horizon and decreasing to negative values. The zero-crossing points, marked as extrema, show the unstable photon sphere. For smaller $r_+$, crossings occur at smaller $r$ with more extreme negative slopes post-crossing, signifying sharper potential peaks. Larger $r_+$ shift crossings outward and moderate the slopes, consistent with weakened instability. As $r_+$ grows further, the derivative may fail to cross zero, confirming the disappearance of extrema and thus no Lyapunov exponent for those configurations.

\begin{figure}[htbp]
\centering
\begin{subfigure}{0.45\textwidth}
\includegraphics[width=\textwidth]{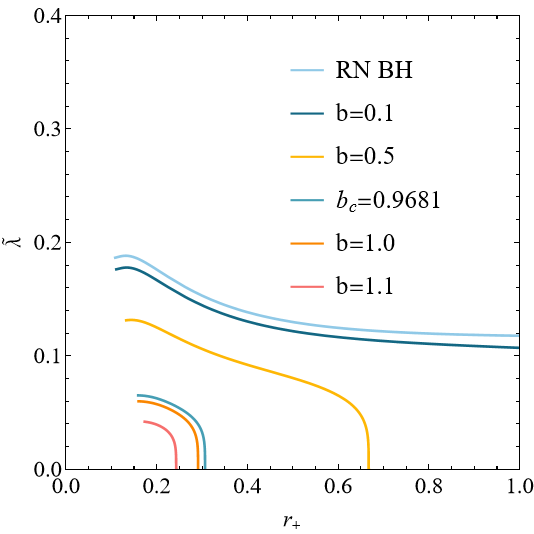}
\caption{$\tilde{\lambda} - r_+$ for various b. }
\label{fig:lambda_r+(null)}
\end{subfigure}
\hfill
\begin{subfigure}{0.45\textwidth}
\includegraphics[width=\textwidth]{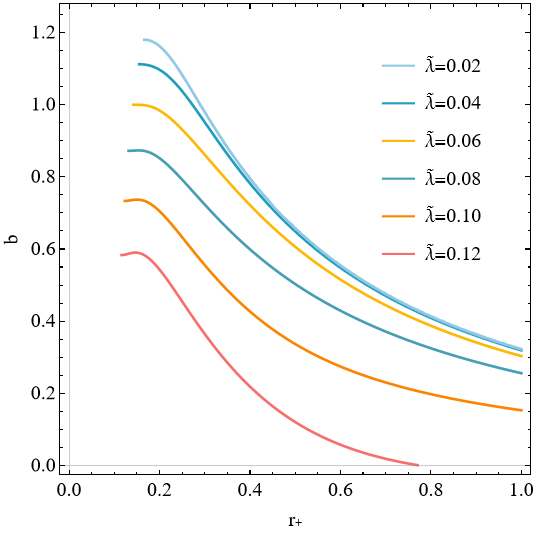}
\caption{$\tilde{\lambda}\left(r_{+}, b\right)$. }
\label{fig:lambda_r+b(null)}
\end{subfigure}
\caption{Lyapunov exponent $\lambda$ of massless particles on unstable null circular orbits log-scaled in contour as $\tilde{\lambda} = \log_{100}(\lambda + 1)$ for horizon radius $r_+$ and quintessence normalization factor $b$, with $T(r_+, b) > 0$. }
\label{fig:lambda_analysis(null)}
\end{figure}

The instability of circular orbits is evaluated using the Lyapunov exponent, where positive values indicate chaotic dynamics. We examine the Lyapunov exponent $\lambda$ as a function of event horizon radius $r_+$ and the quintessence normalization factor $b$. The relations, with other spacetime and particle parameters held constant, are presented in Fig.~\ref{fig:lambda_analysis(null)}.
In Fig.~\ref{fig:lambda_r+(null)}, Lyapunov exponent $\lambda$ decreasing with $r_+$ for various $b$ values. these functions exhibit a left-side cutoff as a result of the need for a positive Hawking temperature $T(r_+, b)>0$. Unlike in RN black hole, where unstable circular orbits always exist, higher $b$ lowers $\lambda$ overall, with curves approaching zero at large $r_+$. This corresponds to the disappearance of unstable circular orbits, consistent with the discussion in Fig.~\ref{fig:Veff_analysis(null)}. It indicates the reduced chaos for larger horizons and stronger quintessence effects. Moreover, The contour plot in Figure~\ref{fig:lambda_r+b(null)} maps the Lyapunov exponent $\lambda$ across the $r_+-b$ space, revealing distinct regimes influenced by the quintessence parameter $b$. For small $b$ below $1.2$, $\lambda$ exists and exhibits non-zero values, starting high at small $r_+$ (indicating strong chaos near the unstable photon sphere) and decreasing monotonically as $r_+$ increases, eventually approaching zero due to the flattening of the effective potential and disappearance of unstable circular orbits. In contrast, for larger $b$ exceeding $1.2$, $\lambda$ vanishes entirely. Overall, increasing $b$ systematically suppresses $\lambda$, demonstrating quintessence's effects in restrain orbital chaos in RN-AdS black holes.

\begin{figure}[htbp]
\centering
\begin{subfigure}{0.47\textwidth}
\includegraphics[width=\textwidth]{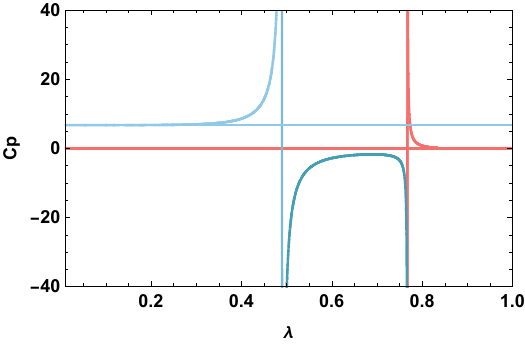}
\caption{$C_p$ for $\lambda$. }
\label{fig:Cp_vs_lambda(null)}
\end{subfigure}
\hfill
\begin{subfigure}{0.47\textwidth}
\includegraphics[width=\textwidth]{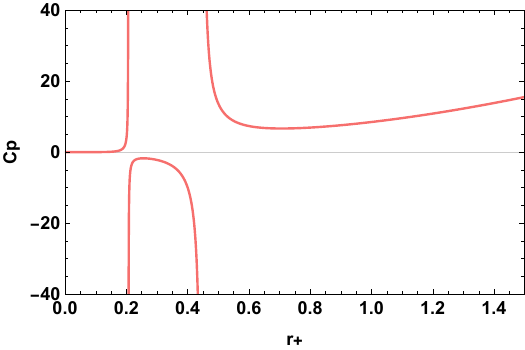}
\caption{$C_p$ for $r_+$. }
\label{fig:Cp_vs_r_+}
\end{subfigure}
\caption{Isobaric heat capacity $C_p$ for $\lambda$ and $r_+$, delineating stability regions: positive $C_p$ in SBH/LBH (stable, varying $\lambda$), negative in IBH (unstable, high $\lambda$) and $b=0.5 < b_c$.}
\label{fig:Cp_analysis(null)}
\end{figure}

We now turn to the isobaric heat capacity 
\begin{align}
    C_p = -T^2 \partial^2 F / \partial T^2
    \label{eq:cp}
\end{align}
with Eq. \eqref{eq:T} and Eq. \eqref{eq:F}. The heat capacity  as functions of Lyapunov exponent and event horizon radius for cases exhibiting phase transitions with $b=0.5 < b_c$ are shown in Fig.~\ref{fig:Cp_analysis(null)}.
In Fig.~\ref{fig:Cp_vs_lambda(null)} which the case for $b$ below the critical value, the heat capacity $C_p$ is positive when the Lyapunov exponent $\lambda$ is either large or small, but becomes negative in the intermediate $\lambda$ region. This indicates that both small and large black holes remain stable, while intermediate black holes exhibit instability and enhanced chaotic behavior.
Similarly, where heat capacity remains positive at both small and large $r_+$ but turns negative in between. This again demonstrates that black holes at the extremes are stable, whereas those in the intermediate regime are unstable and more chaotic.  

\begin{figure}[htbp]
\centering
\begin{subfigure}{0.47\textwidth}
\includegraphics[width=\textwidth]{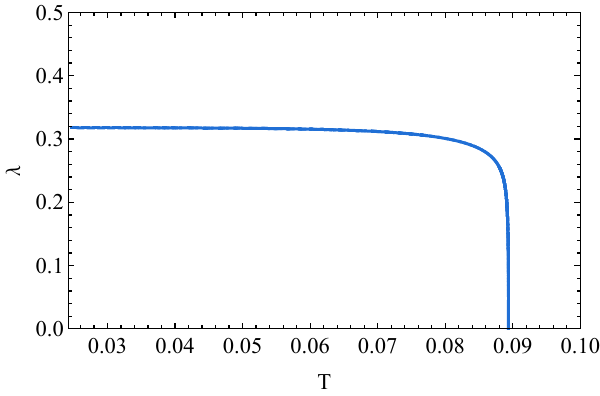}
\caption{ $b=1 > b_c$.}
\label{fig:lambda_T_b1_null}
\end{subfigure}
\hfill
\begin{subfigure}{0.47\textwidth}
\includegraphics[width=\textwidth]{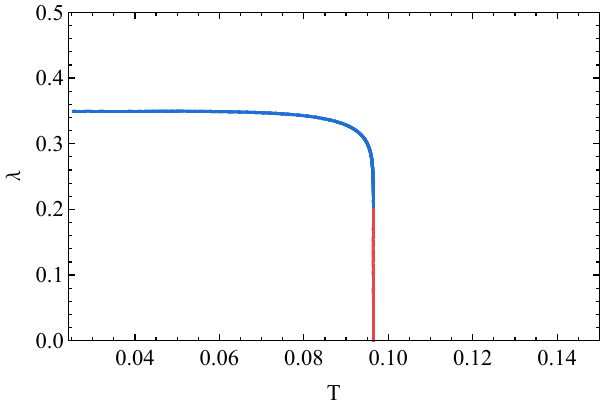}
\caption{ $b = b_c=0.9681$.}
\label{fig:lambda_T_bc_null}
\end{subfigure}

\vspace{1cm}

\begin{subfigure}{0.47\textwidth}
\centering
\includegraphics[width=\textwidth]{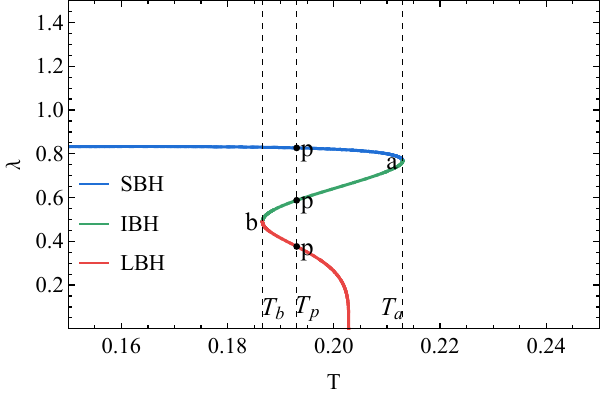}
\caption{$b=0.5 < b_c$.}
\label{fig:lambda_T_b05_null}
\end{subfigure}
\hfill
\begin{subfigure}{0.47\textwidth}
\centering
\includegraphics[width=\textwidth]{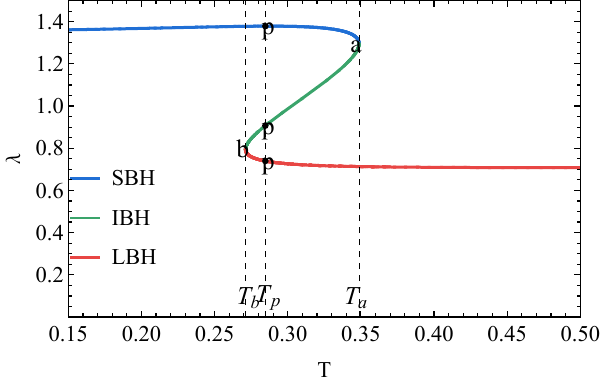}
\caption{ $b=0 < b_c$.}
\label{fig:lambda_T_b0_null}
\end{subfigure}
\caption{Lyapunov exponent $\lambda$ with Hawking temperature $T$ for null geodesics with varying $b$. Multivalued branches  emerge for $b < b_c$, single-valued curves appear for $b > b_c$. Fixed $Q=0.1$}
\label{fig:lambda_T_null}
\end{figure}

Furthermore, we explore the relationship between Lyapunov exponent $\lambda$ and Hawking temperature $T$ in Fig.~\ref{fig:lambda_T_null} for varying $b$ with $Q=0.1$. For $b=1 > b_c$, Fig.~\ref{fig:lambda_T_b1_null} illustrates $\lambda$ decreasing monotonically with $T$ and approaching zero at a finite temperature cutoff, beyond which no unstable null circular orbits exist, indicating the absence of chaotic photon dynamics in this regime. At the critical $b = b_c$, Fig.~\ref{fig:lambda_T_bc_null} reveals an inflection point that signals of criticality, and again terminating at zero for sufficiently high $T$. In contrast, for $b < b_c$ such as $b=0.5$ and $b=0$, Fig.s~\ref{fig:lambda_T_b05_null} and \ref{fig:lambda_T_b0_null} display pronounced multivalued behavior within the temperature interval $T_b < T < T_a$, where three distinct branches coexist: the small black hole (SBH) branch at low $T$ with high $\lambda$ reflecting strong instability, the intermediate black hole (IBH) branch with transitional $\lambda$ values, and the large black hole (LBH) branch at high $T$ with low $\lambda$ approaching zero. At the phase transition temperature $T_p$, $\lambda$ undergoes a discontinuous jump from the SBH to LBH branch, characteristic of a first-order transition driven by the equal free energy condition. A key distinction emerges when comparing to the pure RN case at $b=0$. while RN black holes maintain a persistent non-zero asymptotic $\lambda$ at high $T$ or large $r_+$, the quintessence ($b \neq 0$) imposes a finite cutoff where $\lambda$ vanishes entirely. This situation similar to behaviors in massive particle geodesics but novel for null orbits. Compare with Fig.~\ref{fig:FvsT_b_various}, the multivalued $\lambda-T$ structures precisely correspond to the swallow tail configurations in free energy, where $\lambda$ discontinuities map onto free energy crossings, forming a direct link between dynamical chaos and thermodynamic phases. 
Quintessence reducing $\lambda$ with increasing $b$ but also brings a cut off not found in RN-AdS spacetimes, providing a dynamical signature for probing phase transitions and influence of dark energy on null geodesic instability.

\begin{figure}[htbp]
\centering
\begin{subfigure}{0.47\textwidth}
\includegraphics[width=\textwidth]{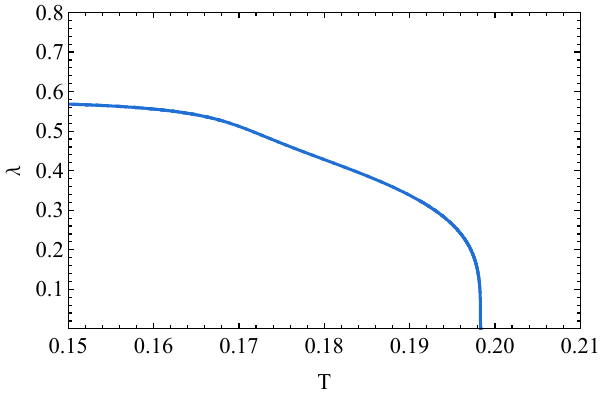}
\caption{$Q=0.15>Q_c$.}
\label{fig:lambda_T_Q0.15_null}
\end{subfigure}
\hfill
\begin{subfigure}{0.47\textwidth}
\includegraphics[width=\textwidth]{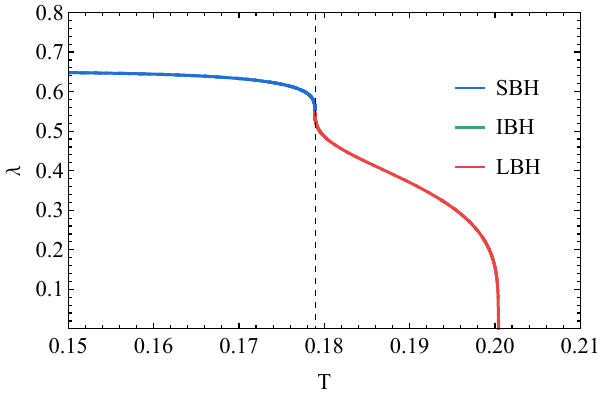}
\caption{$Q=Q_c=0.1295$.}
\label{fig:lambda_T_Qc_null}
\end{subfigure}

\vspace{1cm}

\begin{subfigure}{0.47\textwidth}
\centering
\includegraphics[width=\textwidth]{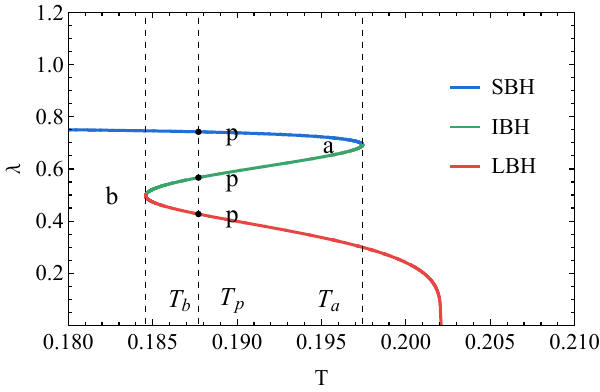}
\caption{$Q=0.11<Q_c$.}
\label{fig:lambda_T_Q0.11_null}
\end{subfigure}
\hfill
\begin{subfigure}{0.47\textwidth}
\centering
\includegraphics[width=\textwidth]{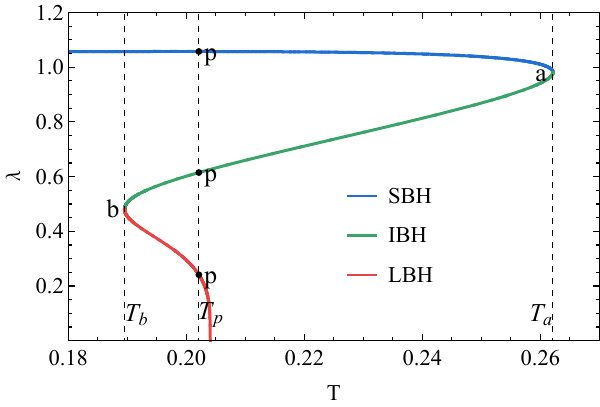}
\caption{$Q=0.08<Q_c$.}
\label{fig:lambda_T_Q0.08_null}
\end{subfigure}
\caption{The Lyapunov exponent $\lambda$ as a function of Hawking temperature $T$ for null geodesics with normalization factor $b=0.5$. For $Q < Q_c$, the Lyapunov exponent exhibits multivalued branches with discontinuities at points indicative of a first-order phase transition, while for $Q > Q_c$, it remains single-valued, reflecting a continuous behavior. }
\label{fig:lambda_T_Q_null}
\end{figure}

The black hole charge $Q$ reveals analogous influences on chaotic dynamics and thermodynamics to dark energy. We further examine the Lyapunov exponent $\lambda$ as a function of Hawking temperature $T$ for null geodesics with varying charge $Q$, as depicted in Fig.~\ref{fig:lambda_T_Q_null}.
For $Q=0.15 > Q_c$, $\lambda$ is observed to decrease monotonically from high values to zero at a finite $T$ in Fig.~\ref{fig:lambda_T_Q0.15_null}, indicating no phase multiplicity and a limited chaotic regime.
At the critical charge $Q_c = 0.1295$, a steep decline with an inflection point emerges in Fig.~\ref{fig:lambda_T_Qc_null}, marking a criticality.
For $Q < Q_c$, such as $Q=0.11$ and $Q=0.08$, multivalued behavior is revealed in the range $T_b < T < T_a$ through Figs.~\ref{fig:lambda_T_Q0.11_null} and \ref{fig:lambda_T_Q0.08_null}, featuring three branches: small black hole (SBH, blue, high $\lambda$), intermediate black hole (IBH, green, transitional $\lambda$), and large black hole (LBH, red, low $\lambda$ approaching zero).
At the transition temperature $T_p$, $\lambda$ exhibits a discontinuous jump from SBH to LBH, signifying a first-order phase transition aligned with free energy equality.
This multivalued $\lambda-T$ structure mirrors the swallowtail in free energy landscapes (cf. Fig.~\ref{fig:FvsT_b_various}), linking geodesic instability directly to thermodynamic phases.
Increasing charge narrows the temperature range and suppresses multivalued branches, while quintessence ensures $\lambda$ vanishes at high $T$, offering a dynamical probe of charge effects on massless particles in unstable circular orbits in dark energy-modified spacetimes.

\subsection{Lyapunov Exponents and Thermodynamics for Timelike Geodesics}
\label{sub:timelike}

\begin{figure}[htbp]
\centering
\begin{subfigure}{0.47\textwidth}
\includegraphics[width=\textwidth]{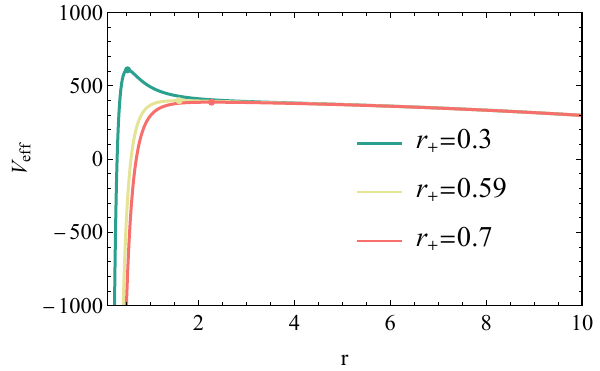}
\caption{$V_{\text{eff}}-r$ for varying $r_+$.}
\label{fig:Veff_time}
\end{subfigure}
\hfill
\begin{subfigure}{0.47\textwidth}
\includegraphics[width=\textwidth]{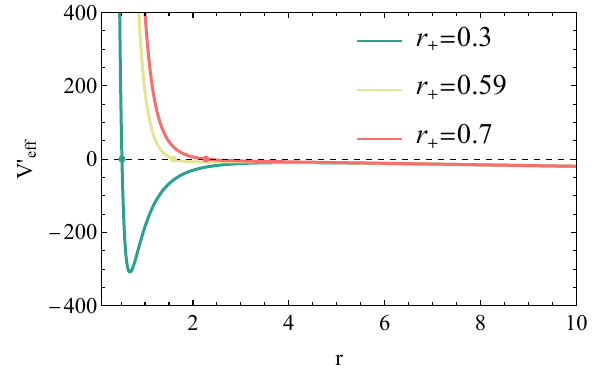}
\caption{$V_{\text{eff}}'-r$ for varying $r_+$.}
\label{fig:Veff_prime_time}
\end{subfigure}
\caption{Effective potential $V_{\text{eff}}(r)$ and its derivative for different $r_+$, with $Q=0.1$. }
\label{fig:Veff_analysis}
\end{figure}

Building on the thermodynamic analysis from previous sections, we now explore the connection between the Lyapunov exponent for timelike geodesics and the thermodynamics of RN-qAdS black holes. Massive particles move along timelike geodesics in the equatorial plane ($\theta=\pi / 2$), described by the Hamiltonian
\begin{align}
- 2 \mathcal{H} = f(r) \dot{t}^2 - \frac{\dot{r}^2}{f(r)} - r^2 \dot{\varphi}^2 = 1.
\end{align}
 Conserved quantities include energy $E = f(r) \dot{t}$ and angular momentum $L = r^2 \dot{\varphi}$. The radial equation becomes $\dot{r}^2 + V_{\text{eff}}(r) = E^2$, where the effective potential is
\begin{align}
    V_{\text{eff}}(r) = f(r) \left( 1 + \frac{L^2}{r^2} -\frac{E^2}{f(r)} \right),
\end{align}
where $L=\tilde{L}/l$ is scaled with angular momentum. The unstable circular orbits satisfy $V_{\text{eff}}'(r_c) = 0$ and $V_{\text{eff}}''(r_c) < 0$. The effective potential  $V_{\text{eff}}$  and its derivative  $V_{\text{eff}}^\prime$  with respect to $r$ for various values of $ r_+$ are shown in Fig.~\ref{fig:Veff_analysis}. Similar to the massless particle scenario, The effective potential $ V_{\text{eff}}$ first increases and then decreases with $r$. Smaller $r_+$ yield higher maxima at larger $r$, indicating stronger barriers for unstable orbits. Larger $r_+$ flatten the potential, reducing peak heights and shifting them outward, which suggests decreased instability. The derivative of the effective potential with respect to $r$ exhibits in Fig.~\ref{fig:Veff_prime_time}, a local minimum exist for smaller values of $r_+$.

\begin{figure}[htbp]
\centering
\begin{subfigure}{0.45\textwidth}
\includegraphics[width=\textwidth]{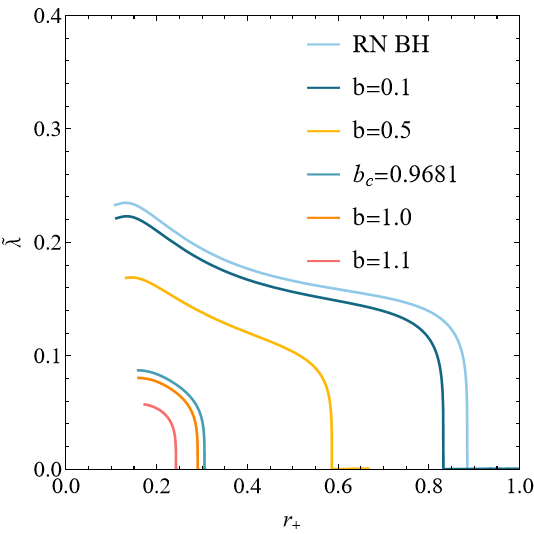}
\caption{$\tilde{\lambda}-r_+$ for various b.}
\label{fig:lambda_r+(timelike)}
\end{subfigure}
\hfill
\begin{subfigure}{0.45\textwidth}
\includegraphics[width=\textwidth]{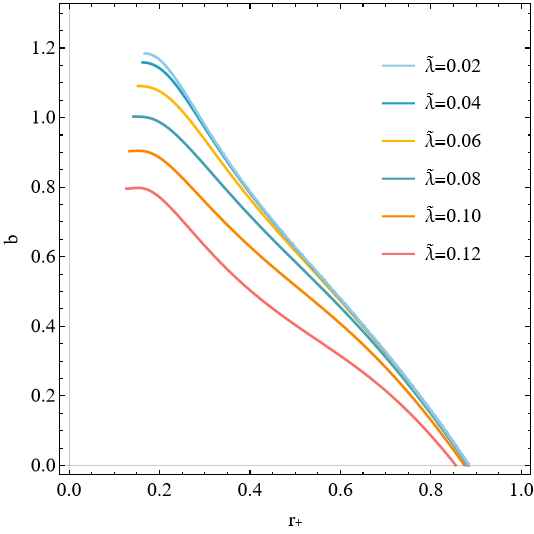}
\caption{$\tilde{\lambda}\left(r_{+}, b\right)$.}
\label{fig:lambda_r+b(timelike)}
\end{subfigure}
\caption{Lyapunov exponent $\lambda$ of massive particles on unstable null circular orbits log-scaled in contour as $\tilde{\lambda} = \log_{100}(\lambda + 1)$.}
\label{fig:lambda_analysis(timelike)}
\end{figure}

\begin{figure}[htbp]
\centering
\begin{subfigure}{0.47\textwidth}
\includegraphics[width=\textwidth]{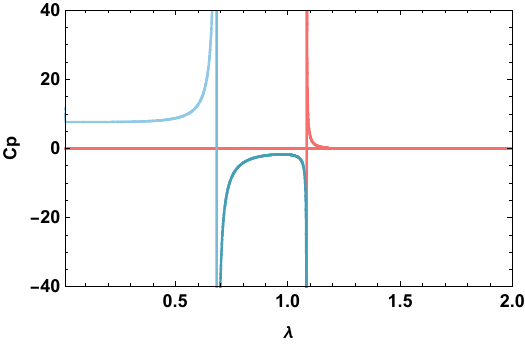}
\caption{$C_p-\lambda$ for $b=0.5<b_c$.}
\label{fig:Cp_vs_lambda_time}
\end{subfigure}
\hfill
\begin{subfigure}{0.47\textwidth}
\includegraphics[width=\textwidth]{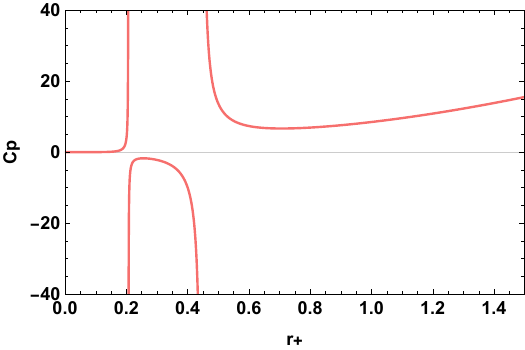}
\caption{$C_p-r_+$ for $b=0.5<b_c$.}
\label{fig:Cp_vs_r+time}
\end{subfigure}
\caption{Isobaric heat capacity $C_p$ versus $\lambda$ and $r_+$, highlighting stability: positive $C_p$ in SBH/LBH stable with high-to-low $\lambda$, negative in IBH unstable with transitional $\lambda$.}
\label{fig:Cp_lambda_analysis}
\end{figure}

The local maxima and minima of the effective potential determine the stable or unstable equilibria of a particle and can be linked to the Lyapunov exponent. The Lyapunov exponent for these orbits is \cite{Lyu:2023sih}
\begin{equation}
\lambda = \frac{1}{2} \sqrt{-\frac{r_c^3 f'(r_c) V_{\text{eff}}''(r_c)}{L^2}}, 
\end{equation}
where positive values indicate instability and quantify chaos. 
 Lyapunov exponent quantifies the exponential divergence rate of neighboring trajectories in a dynamical system and serves as a fundamental parameter for characterizing chaotic dynamics. From the above equation, the dependence of the Lyapunov exponent $\lambda$ on $r_{+}$ and the impact parameter $b$ is obtained, the results for charge $Q=0.1$ and angular momentum $L=20$ are displayed in Fig.~\ref{fig:lambda_analysis(timelike)}.
In Fig.~\ref{fig:lambda_r+(timelike)}, Lyapunov exponent $\lambda$ decreases with $r_+$ for all $b$, starting high at small event horizons with the condition $T(r_+, b)>0$ and approaching zero at finite horizons radii. Larger $b$ suppress $\lambda$ overall, with the RN black hole case $b=0$ showing the highest values. This demonstrates reduced chaos for bigger black holes and stronger quintessence. The contour in Fig.~\ref{fig:lambda_r+b(timelike)} features the high Lyapunov exponent $\lambda$ at small $r_+$ and low $b$, as $r_+$ or $b$ grows, the $\lambda$ drops to zero because the unstable equilibrium vanishes. We also use Eq. \eqref{eq:cp} to plot $C_p$ for $b=0.5<b_c$ in Fig.~\ref{fig:Cp_lambda_analysis}.  heat capacity remains positive at low and high $\lambda$ or $r_+$, negative intermediately, corresponding to stable SBH/LBH with varying chaos and unstable IBH with peak instability.

\begin{figure}[htbp]
\centering
\begin{subfigure}{0.47\textwidth}
\includegraphics[width=\textwidth]{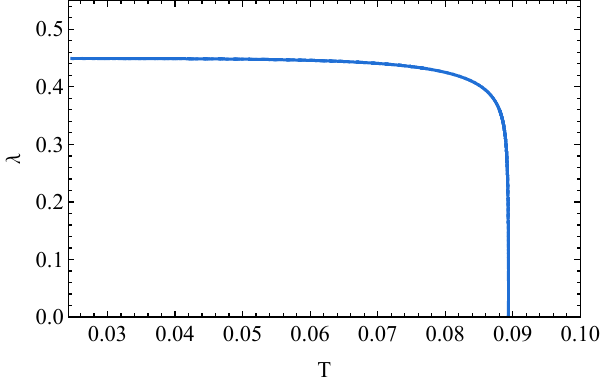}
\caption{$b=1>b_c$.}
\label{fig:lambda_T_b1_time}
\end{subfigure}
\hfill
\begin{subfigure}{0.47\textwidth}
\includegraphics[width=\textwidth]{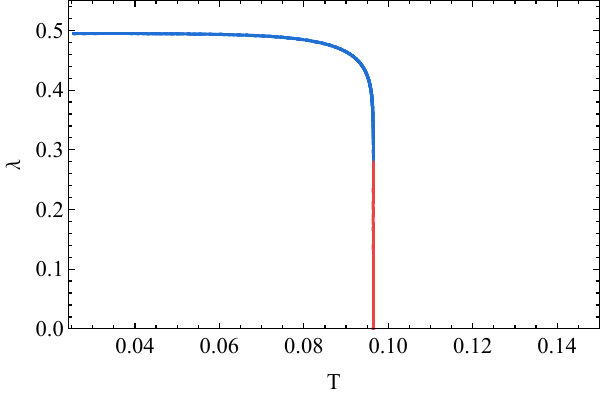}
\caption{$b=b_c=0.9681$.}
\label{fig:lambda_T_bc_time}
\end{subfigure}

\vspace{1cm}

\begin{subfigure}{0.47\textwidth}
\centering
\includegraphics[width=\textwidth]{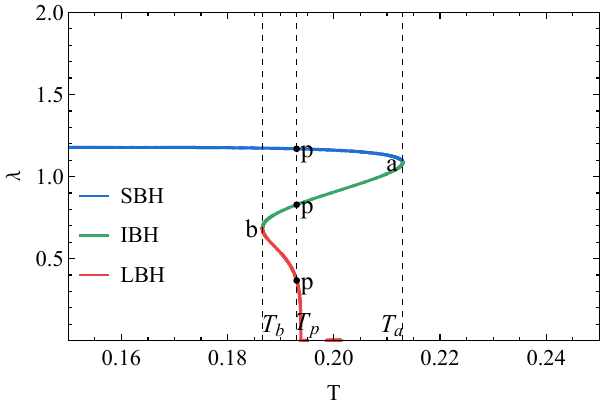}
\caption{$b=0.5<b_c$.}
\label{fig:lambda_T_b05_time}
\end{subfigure}
\hfill
\begin{subfigure}{0.47\textwidth}
\centering
\includegraphics[width=\textwidth]{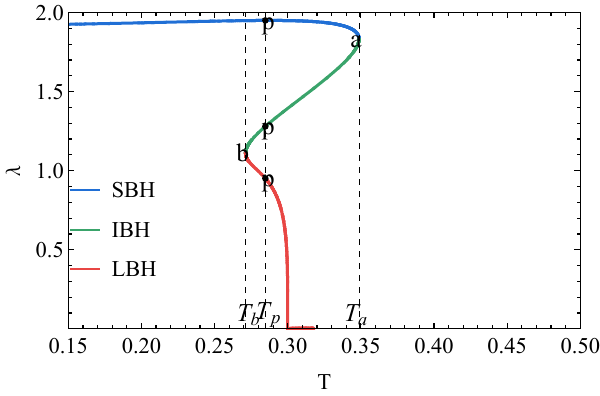}
\caption{$b=0<b_c$.}
\label{fig:lambda_T_b0_time}
\end{subfigure}
\caption{Lyapunov exponent $\lambda$ versus Hawking temperature $T$ for timelike geodesics. Multivalued branches appear for $b<b_c$, with jumps at transition points reflecting first-order shifts; single-valued for $b>b_c$.}
\label{fig:lambda_T_timelike}
\end{figure}

We have shown $\lambda$ as a function of $T$ for different $b$ with other parameters fixed in Fig.~\ref{fig:lambda_T_timelike} . For $b=1>b_c$, $\lambda$ decreases steadily with $T$, denoting a single phase. At $b=b_c$, an inflection point emerges, indicating criticality. For $b=0.5<b_c$ and $b=0<b_c$, multivalued structures form SBH branch at low $T$ with high $\lambda$, IBH with transitional values, LBH at high $T$ with low $\lambda$. Discontinuous jumps at $T_p$ align with equal free energies, signaling first-order transitions. These patterns correlate with free energy swallowtails, linking dynamical instability to thermodynamic phases. Even though quintessence suppresses chaos, the Lyapunov exponent of unstable circular orbits for photons and massive particles can still serve as a tool to investigate phase transitions.

\begin{figure}[htbp]
\centering
\begin{subfigure}{0.47\textwidth}
\includegraphics[width=\textwidth]{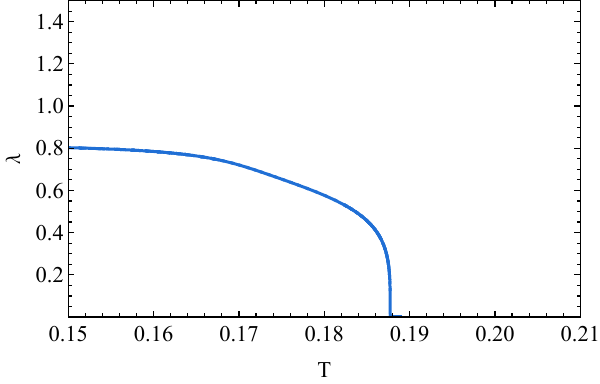}
\caption{$Q=0.15>Q_c$.}
\label{fig:lambda_T_Q0.15_time}
\end{subfigure}
\hfill
\begin{subfigure}{0.47\textwidth}
\includegraphics[width=\textwidth]{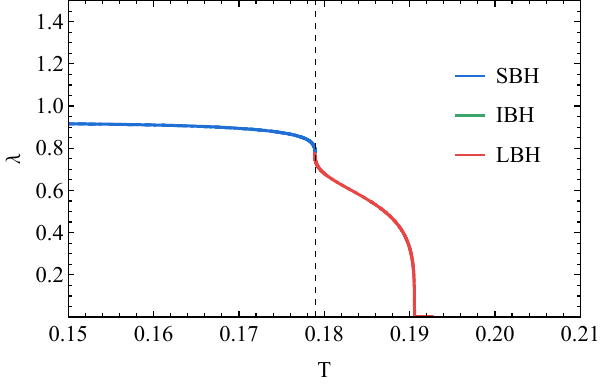}
\caption{$Q=Q_c=0.1295$.}
\label{fig:lambda_T_Qc_time}
\end{subfigure}

\vspace{1cm}

\begin{subfigure}{0.47\textwidth}
\centering
\includegraphics[width=\textwidth]{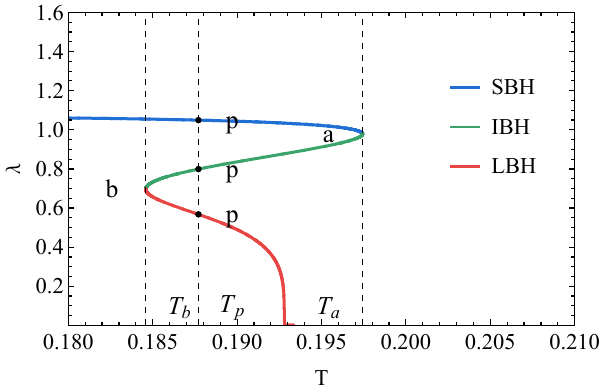}
\caption{$Q=0.11<Q_c$.}
\label{fig:lambda_T_Q0.11_time}
\end{subfigure}
\hfill
\begin{subfigure}{0.47\textwidth}
\centering
\includegraphics[width=\textwidth]{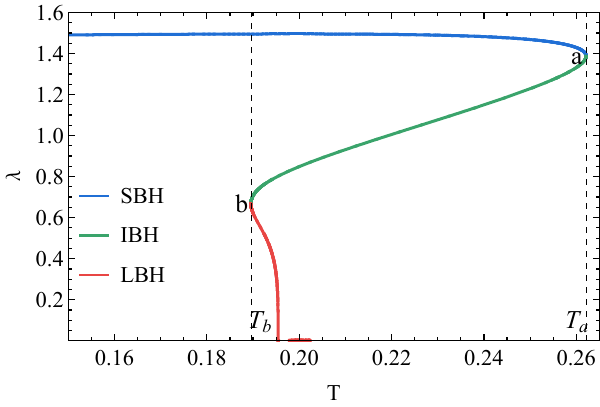}
\caption{$Q=0.08<Q_c$.}
\label{fig:lambda_T_Q0.08_time}
\end{subfigure}
\caption{Lyapunov exponent $\lambda$ versus Hawking temperature $T$ for timelike geodesics.}
\label{fig:lambda_T_Q_timelike}
\end{figure}

We further investigate the role of electric charge by varying Q while fixing other parameters, revealing similar dynamical signatures in the Lyapunov exponent. In Fig.~\ref{fig:lambda_T_Q_timelike}, $\lambda$ decreases monotonically with T for $Q=0.15 > Q_c$, indicating a single stable phase. At $Q=Q_c=0.1295$, an inflection point appears, marking the critical boundary. For $Q=0.11< Q_c$ and$ Q=0.08 < Q_c$, multivalued branches emerge: SBH at low T with high $\lambda$, IBH in the intermediate range, and LBH at high $T$ with low $\lambda$. Discontinuities arise at transition temperatures where free energies are equal, indicating first-order phase transitions. These behaviors reflect the thermodynamic swallowtail structures, confirming that variations in charge modulate chaotic dynamics while maintaining the Lyapunov exponent as an effective probe for phase transitions.

\section{The Critical Exponent of RN-qAdS black hole with Lyapunov Exponent}
\label{sec:critical_exponent}

To probe the critical behavior of RN-qAdS black hole near phase transition points, we can use the discontinuous change of the Lyapunov exponents $\lambda$ at the first-order phase transition point $p$ in Fig.~\ref{fig:lambda_T_timelike} and Fig.~\ref{fig:lambda_T_null} to study the critical behavior of black holes. Denoting $\lambda_s$ for the small black hole (SBH) and $\lambda_l$ for the large black hole (LBH), we define the order parameter as
\begin{align}
    \Delta \lambda = \lambda_l - \lambda_s.
\end{align}
At the critical temperature $T_c$, we have $\lambda_s = \lambda_l = \lambda_c$, yielding $\Delta \lambda = 0$. This vanishing at criticality positions $\Delta \lambda$ as a robust probe for phase transition dynamics.

The critical exponent $\delta$ governs the scaling of $\Delta \lambda$ near $T_c$
\begin{align}
    \Delta \lambda \sim |T - T_c|^\delta.
\end{align}
To determine $\delta$, we employ a Taylor expansion of $\lambda$ around the critical point:
\begin{align}
    \lambda = \lambda_c + \left[ \frac{\partial \lambda}{\partial r_+} \right]_c (r_+ - r_{+,c}) + \mathcal{O}(r_+ - r_{+,c})^2
\end{align}
where the subscript $c$ indicates critical values. At the phase transition, let $r_p = r_{+,c} (1 + \Delta)$ with $|\Delta| \ll 1$. For SBH and LBH, offsets $\Delta_s$ and $\Delta_l$ give
\begin{align}
    \Delta \lambda = \lambda_l - \lambda_s \approx \left[ \frac{\partial \lambda}{\partial r_+} \right]_c (r_l - r_s),
\end{align}
where $r_l = r_{+,c} (1 + \Delta_l)$ and $r_s = r_{+,c} (1 + \Delta_s)$. Relating $r_l - r_s$ to temperature requires expanding $T(r_+)$ with 
\begin{align}
    T(r_+) = T_c + \frac{1}{2} \left[ \partial^2 T/\partial r_+^2\right]_c (r_+ - r_{+,c})^2 + \mathcal{O}(r_+ - r_{+,c})^3.
\end{align}
At transition, $T(r_s) = T(r_l) = T_p$. Setting $T_p = T_c (1 + \epsilon)$ with $|\epsilon| \ll 1$, we find
\begin{align}
    r_l - r_s = 2 \sqrt{ \frac{2 T_c \epsilon}{\left[ \partial^2 T/\partial r_+^2 \right]_c} }.
\end{align}
Substituting yields
\begin{align}
    \Delta \lambda \approx \left[ \frac{\partial \lambda}{\partial r_+} \right]_c \cdot 2 \sqrt{ \frac{2 T_c \epsilon}{\left[ \partial^2 T/\partial r_+^2 \right]_c} }.
\end{align}
With $\epsilon = (T_p - T_c)/T_c$, as $T_p \to T_c$,
\begin{align}
    \Delta \lambda \sim |T_p - T_c|^{1/2},
\end{align}
implying $\delta = 1/2$. 

\begin{figure}[htbp]
    \centering
    \begin{subfigure}{0.46\textwidth}
        \includegraphics[width=\textwidth]{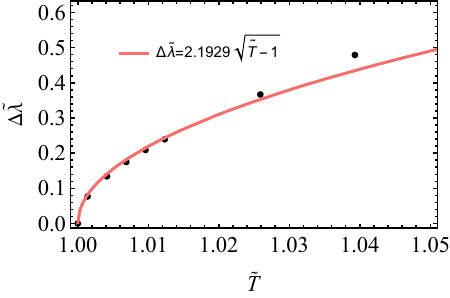}
        \caption{Timelike geodesics.}
        \label{fig:delta_lambda(timelike)_T}
    \end{subfigure}
    \hfill
    \begin{subfigure}{0.46\textwidth}
        \includegraphics[width=\textwidth]{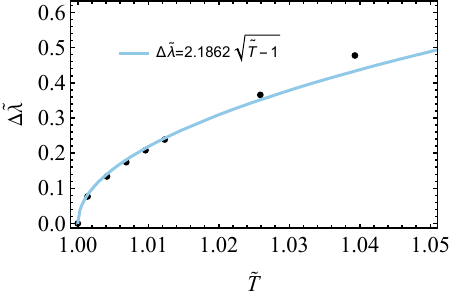}
        \caption{Null geodesics.}
        \label{fig:delta_lambda(null)_T}
    \end{subfigure}
    \caption{Lyapunov exponents plotted $\tilde{\lambda}=\Delta \lambda / \lambda_c$ as a function of the rescaled phase transition temperature $\tilde{T}=T_p / T_c$.}
    \label{fig:delta_lambda_T}
\end{figure}

In the quintessence context, we compute $\Delta\tilde{\lambda}=\Delta \lambda / \lambda_c$ as a function for $\tilde{T}=T_p / T_c$ by choosing points which near the critical point in Fig.~\ref{fig:b_vs_q}, plotting results for timelike and null geodesics in Fig.~\ref{fig:delta_lambda_T}.
As shown in Fig.~\ref{fig:delta_lambda_T}, $\Delta\tilde{\lambda}$ approaches zero as $T_p$ nears $T_c$. Numerical fits over $\tilde{T}$ yield 
\begin{align}
    \Delta \tilde{\lambda} = k \sqrt{\tilde{T} - 1}, 
\end{align}
aligning with $\delta = 1/2$. This analysis reveals that $\Delta \lambda$ acts as an order parameter with critical exponent $1/2$ for charged AdS black holes in quintessence, consistent with studies on Gauss-Bonnet, RN-AdS, and Born-Infeld AdS black holes \cite{Lyu:2023sih, Guo:2022kio, Yang:2023hci}.

\section{Conclusion and Discussion}
\label{sec:concl}
In this paper, we have investigated the thermodynamic phase transitions of Reissner-Nordström-AdS (RN-AdS) black holes surrounded by quintessence through Lyapunov exponents. By incorporating the quintessence field, we generalized the RN-AdS framework to include dark energy effects, revealing a modified parameter space in the $b-Q$ plane that delineates regions with van der Waals-like small/large black hole phase transitions. Numerical analysis of temperature-horizon radius relations and free energy landscapes confirmed the presence of first-order transitions below criticality and second-order transitions at the critical point, with quintessence introducing an upper bound for phase transitions to occur.

To explore the dynamical signatures of these thermodynamic phases, we computed Lyapunov exponents for both null and timelike geodesics in unstable circular orbits near the event horizon. For massless particles (null geodesics), the effective potential analysis showed that increasing the quintessence parameter $b$ or horizon radius $r_+$ suppresses instability, leading to a finite cutoff where unstable orbits vanish—a feature absent in pure RN-AdS spacetimes. Similarly, for massive particles (timelike geodesics), Lyapunov exponents decrease with $r_+$ and $b$, approaching zero at large horizons. Contour plots of $\lambda$ as a function of $r_+$ and $b$ highlighted quintessence's role in reducing chaos, while $C_p-\lambda$ diagrams linked positive heat capacity (stable small/large black holes) to varying $\lambda$ levels and negative heat capacity (unstable intermediate black holes) to transitional instability.

The relationship between Lyapunov exponents and Hawking temperature further illuminated the phase structure. For $b < b_c$ or $Q < Q_c$, $\lambda-T$ diagrams exhibit multivalued branches corresponding to small, intermediate, and large black holes, with discontinuous jumps at the first-order transition temperature $T_p$ mirroring free energy swallowtails. At $b = b_c$ or $Q = Q_c$ , an inflection point signals second-order criticality, and for $b > b_c$ or $Q > Q_c$, monotonic single-valued behavior indicates no transitions. Notably, quintessence imposes a temperature cutoff where $\lambda$ vanishes for both geodesic types, distinguishing it from RN-AdS cases where null geodesic exponents persist asymptotically. we also find the discontinuity $\Delta \lambda = \lambda_l - \lambda_s$ at $T_p$ serves as an order parameter, yielding a critical exponent of $1/2$ consistent with van der Waals systems.

Our findings demonstrate that Lyapunov exponents provide a powerful dynamical probe for thermodynamic phase transitions in quintessence black holes, offering insights into dark energy's influence on black hole dynamics and thermodynamics. Future work could extend this to higher-dimensional quintessence models.

\begin{acknowledgments}
The authors express their gratitude to Guangzhou Guo, Yuling Weng and Yang Cao for their valuable suggestions and opinions, which have contributed signiffcantly to the completion of this article. This work is supported by the National Natural Science Foundation of China (NSFC) with Grants No. 12175212, and development Fund Project of Shanghai University of Finance and Economics Zhejiang College for the Year 2024 with Grants No. 2024FZJJ02. And it is finished on the server from Kun-Lun in College of Physics, Sichuan University.

\end{acknowledgments}

\normalem

\end{document}